\documentclass[a4paper,11pt]{article}
\usepackage{pos}

\title{Symmetric mass generation and the Nielsen-Ninomiya theorem}
\ShortTitle{Symmetric mass generation}

\author*[a,b]{Maarten Golterman}
\author[c]{Yigal Shamir}

\affiliation[a]{Dept. of Physics and IFAE-BIST, Univ. Aut\`onoma de Barcelona, \\
E-08193 Bellaterra, Barcelona, Spain}

\affiliation[b]{On leave from Dept. of Physics and Astronomy, San Francisco State University,\\
  San Francisco, CA 94132, USA}

\affiliation[c]{Raymond and Beverly Sackler School of Physics and Astronomy,\\
Tel~Aviv University, 69978, Tel~Aviv, Israel}

\emailAdd{maarten@sfsu.edu}
\emailAdd{shamir@tauex.tau.ac.il}

\abstract{The symmetric mass generation (SMG) approach to the
construction of lattice chiral gauge theories attempts to use interactions to render mirror fermions massive without symmetry breaking, to obtain the desired chiral massless
spectrum  (before the gauge field is turned on).  If the zeros that often replace the mirror poles of fermion two-point functions in an SMG phase are ``kinematical'' singularities, general constraints can be formulated on the existence of a chiral fermion spectrum
which are valid in the presence of (non-gauge) interactions of arbitrary strength, including in any SMG phase.
Constructing a one-particle lattice hamiltonian describing the fermion spectrum, we discuss the conditions for the
applicability of the Nielsen-Ninomiya theorem to this hamiltonian. If these conditions are satisfied, the massless fermion spectrum must be vector-like.}

\FullConference{The 42nd International Symposium on Lattice Field Theory (LATTICE2025)\\
2-8 November 2025\\
Tata Institute of Fundamental Research, Mumbai, India\\}

\def\sl#1{\rlap{\hbox{$\mskip 1 mu /$}}#1}
\def\ttl#1{{\it #1}}

\begin{document}
\maketitle

\section{Introduction}
Recently, much work has been done on quantum field theory models in which
``symmetric mass generation'' (SMG) takes place, with an eye toward the construction
of chiral gauge theories on the lattice.    The problem itself as well as attempts to solve
it go back over 40 years.   The problem was first pointed out
in Ref.~\cite{KS}, and formalized in the Nielsen-Ninomiya (NN) theorem
\cite{NN},
which states that, under a number of reasonable assumptions, every right-handed
(massless) fermion is accompanied by a left-handed fermion with the same 
conserved charges.   The theorem was formulated for theories of free fermions,
in the hamiltonian formulation, but holds in the euclidean formulation as well
\cite{LK}.   The SMG approach to circumventing the theorems goes back almost
as much, with the first attempt that of Ref.~\cite{EP}.   More recently, there has been much activity, and we refer to the review of Ref.~\cite{WY}
and our own paper \cite{GSNN}, of which these proceedings give a brief 
overview, for detailed lists of references.

The SMG framework can be summarized as follows.  We start with a set of 
free Dirac fermions in
complex representations of a Lie group G, which is the symmetry group to be
gauged.   We can assume that our goal is a chiral gauge theory in
which only left-handed (LH) Weyl fermions occur.\footnote{In 4 dimensions
RH fermions can always be written as LH fermions through charge
conjugation.}

Next, we introduce a (non-gauge) ``strong'' interaction $H_{\rm int}$ that couples
all right-handed (RH) fermion fields; there are no gauge interactions yet.   This defines
the {\it reduced model}.   We then look for a region in the phase diagram of this
strongly coupled theory in which all RH fermions are gapped, while the LH fields
remain massless.   We require that the model has a continuum limit with free 
relativistic massless LH fermions; while the strong interactions are a ``lattice artifact''
solely introduced to give the RH fermions a mass of the order of the lattice
cutoff.   Near the continuum limit the strong interaction is irrelevant, and, in particular, no
spontaneous symmetry breaking (SSB) of G takes place (hence ``symmetric'' mass generation).

If these steps are successful, one can gauge the group G in the final step, and thus construct
a lattice chiral gauge theory.   Of course, this requires the cancellation of 
all (continuous and discrete) gauge anomalies.

\section{The Nielsen-Ninomiya theorem}

\begin{figure}[hbt!]
    \centering
    \vskip-100pt
    \includegraphics[width=0.8\textwidth,angle=-90]{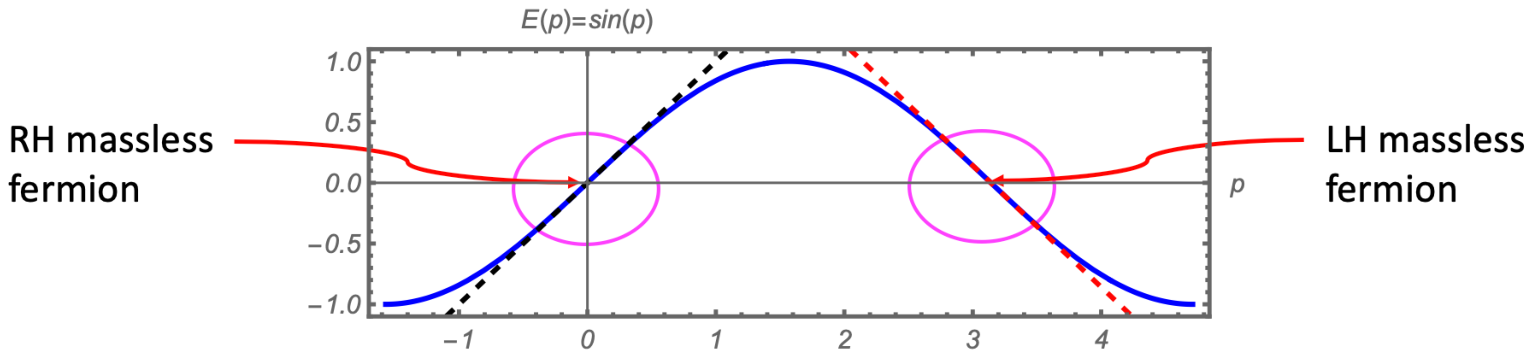}
    \vskip-100pt
    \caption{Illustration of the NN theorem for the dispersion relation $E(p)=\sin(p)$
in one spatial dimension.}
    \label{fig:NN}
\end{figure}
First, we briefly recall the NN theorem.
Consider a free fermionic lattice hamiltonian of the form
\begin{equation}
\label{Hfree}
H=\sum_{xy}\psi^\dagger(x)H(x-y)\psi(y)
\end{equation} 
with the following properties:
\begin{itemize}
\item $H$ is invariant under lattice translations, which implies that momenta
live in a Brillouin zone (a compact manifold, usually a torus).
\item $H(x-y)$ is local, which implies that its eigenvalues $E(\vec p)$ are smooth
functions of momentum.   More precisely, $H(\vec p)$ should have a continuous
first derivative (physically, this allows for the definition of velocity).
\item The massless fermion spectrum in the continuum limit is relativistic.
\item There is an exactly conserved charge with discrete eigenvalues.
\end{itemize}
Under these assumptions, the NN theorem states that for every LH fermion there
exists a RH fermion with the same conserved charge.   As a corollary, if one 
combines this pair into a Dirac fermion, the conserved charge will be vectorlike.
The appearance of these {\it species doublers} ensures that any conserved fermion number symmetry on the lattice will correspond to the fermion number of Dirac fermions in the continuum limit \cite{KS}: species doublers are the tool through which the 
lattice resolves the apparent paradox that the continuum theory has anomalies
in a symmetry which is exactly conserved on the lattice.  In one spatial dimension the situation is 
illustrated in Fig.~\ref{fig:NN} for the dispersion relation $E(p)=\sin(p)$
(setting the lattice spacing $a=1$).

\section{Extension to interacting lattice models.}
The NN theorem does not apply to SMG models, for the simple reason that
it was formulated for free lattice hamiltonians, {\it i.e.}, hamiltonians bilinear 
in the fermion fields with no other fields present, while SMG models are
interacting.   However, this does not mean that the NN theorem cannot be
extended to apply to more complicated (interacting) models as well.   The
key idea \cite{YS93} is to define a lattice effective hamiltonian $H_{\rm eff}$,
that may satisfy the conditions of the NN theorem.   Of course, if it does,
the theorem applies.   It is possible to do so by defining $H_{\rm eff}$ as
the inverse of the retarded propagator at zero frequency ${\cal{R}}(\vec p)$.\footnote{The
reasons for the choice of the retarded propagator are discussed in detail in
Refs.~\cite{GSNN,YS93}.}

What we need are the following requirements:
\begin{itemize}
\item $H_{\rm eff}$ is defined on the same Brillouin zone as the free theory.
This is automatically satisfied.
\item It has to respect the G symmetry with no SSB and well-defined
charge sectors.
\item Only free massless relativistic fermions should appear in the continuum limit;
in particular, no massless bosons should appear.
\item If the underlying lattice model is local, $H_{\rm eff}(\vec p)$ is analytic 
except: (a) a massless fermion corresponds to a pole in ${\cal{R}}(\vec p)$ 
at some momentum, and this leads to a zero in $H_{\rm eff}(\vec p)$ at 
which it is not analytic as we will see below; but it is then 
sufficient to prove that the first derivative is continuous at that zero;
(b) if ${\cal{R}}(\vec p)$ has a zero (a ``propagator zero''), $H_{\rm eff}(\vec p)$ 
has a pole, and the NN theorem simply does not apply.
\end{itemize}
\subsection{Propagator zeros}
Let us first consider the issue of propagator zeros, illustrating this aspect
with a very simple example.   First, decompose the propagator of a massless relativistic
Dirac fermion into its LH and RH parts:\footnote{Here we work in euclidean
space-time.}
\begin{equation}
\label{Diracdecomp}
\frac{i\sl{p}}{p^2}=P_L\,\frac{i\sl{p}}{p^2}\,P_R+P_R\,\frac{i\sl{p}}{p^2}\,P_L\ .
\end{equation}
Here the first term on the right-hand side is our physical Weyl fermion, while 
the second term is the ``mirror'' we wish to gap.   If we are successful at
giving the mirror a mass, the right-hand side changes into
\begin{equation}
\label{mirrorgap}
\longrightarrow P_L\,\frac{i\sl{p}}{p^2}\,P_R+P_R\,\frac{i\sl{p}}{p^2+m^2}\,P_L\ ,
\end{equation}
where $m$ is the mass of the mirror.   But this propagator has a zero!   There are
two possibilities.

First, it is possible that this zero is ``genuine.''   This means that  $H_{\rm eff}(\vec p)$
has a pole, which leads to ghost states and loss of unitarity \cite{Pell,GSzero}.

The second possibility is that the zero is ``kinematical'': the theory contains a LH
partner for the mirror (while the original LH fermion remains massless), and 
Eq.~(\ref{mirrorgap}) changes again, into
\begin{equation}
\label{mirrorgap2}
\longrightarrow P_L\,\frac{i\sl{p}}{p^2}\,P_R+\frac{i\sl{p}+m}{p^2+m^2}
+(?)\,P_R\,\frac{i\sl{p}}{p^2}\,P_L\ .
\end{equation}
The second term now is a massive Dirac fermion, and there is no propagator zero.
The third term, preceded by a question mark, would have to be present if the NN
theorem applies:  a RH massless partner of the massless LH mode would have
to be produced by the theory.   This is precisely the question we wish to
investigate, and we will return to it below.   But first
we discuss how a LH partner for the massive RH mirror could emerge.

\subsection{Bound states}
Since no massive fundamental fields are present in the hamiltonian, the putative
new LH partner of the RH massive mirror would have to emerge as a bound
state.  To illustrate how this may work, it is useful consider a recent example
in one spatial dimension, the 3450 model, a lattice formulation of which was
recently claimed to successfully gap the unwanted mirrors \cite{ZZWY}.

This lattice model contains Dirac fermions $\psi_{3,4,5,0}$ with U(1)
charges 3, 4, 5 and 0, in which the physical fermions correspond to the lattice
fields $\xi_{3,4}=P_R\psi_{3,4}$
and $\xi_{5,0}=P_L\psi_{5,0}$, while the mirrors to 
be gapped correspond to  
$\chi_{3,4}=P_L\psi_{3,4}$
and $\chi_{5,0}=P_R\psi_{5,0}$.\footnote{For the explicit realization of the 
chiral projectors, see Ref.~\cite{ZZWY}.}   If successful,
this would provide a lattice definition of the so-called 3450 U(1) chiral gauge
theory in 2 dimensions.\footnote{In 2 dimensions, LH (RH) Weyl fermions
cannot be mapped onto RH (LH) Weyl fermions.}

To gap the mirrors, Ref.~\cite{ZZWY} introduces an interaction hamiltonian
of the (schematic) form
\begin{equation}
\label{Hint}
H_{\rm int}=g_1\chi_3(\chi^\dagger_4)^2\chi_5\chi_0^2+g_2\chi_3^2\chi_4
(\chi^\dagger_5)^2\chi_0+{\rm h.c.}\ ,
\end{equation}
where squares are to be thought of as point-split.   For an SMG phase to
exist, one expects that the couplings $g_{1,2}$ have to be large.   This
could lead to the generation of bound states with interpolating fields
\begin{equation}
\label{interpolate}
{\cal{B}}_i\propto\frac{\partial H_{\rm int}}{\partial\chi_i^\dagger}\ ,\qquad i=3,\ 4,\ 5,\ 0\ ,
\end{equation}
and the SMG interaction thus leads to mass terms of the form
\begin{equation}
\label{mass}
\chi^\dagger_i{\cal{B}}_i+\mbox{h.c.}
\end{equation}
for each charge,
where each elementary field $\chi_i$
couples to its corresponding bound-state operator ${\cal{B}}_i$.
Kinetic terms for these bound states might be 
induced dynamically as well.
As the only scale in the reduced model is the lattice spacing $a$, one expects 
these masses to be generically of order $1/a$, and this would make the bound states
point-like in the continuum limit.

As discussed in more detail in Ref.~\cite{GSNN}, this mechanism is completely
general, and can happen for any SMG strong interaction!   The interpolating fields
for the putative bound-state partners of the mirror fermions are simply defined
as derivatives of the SMG interaction hamiltonian with respect to each of the
mirror-fermion fields.
In the model of
Ref.~\cite{EP} it was shown that bound-state generation in fact happens in
strong coupling \cite{GPR}.   No analytic strong-coupling expansion is available
in the model of Ref.~\cite{ZZWY}, as the interaction hamiltonian~(\ref{Hint}) is point-split.
Of course, this does not mean that no bound states can be generated, but this
would have to be investigated numerically.\\
\bigskip

Let us get back to the extension of the NN theorem to interacting models.
If the lattice hamiltonian is local, it is unlikely that ghosts associated with
genuine propagator zeros can occur \cite{ZXLY}, so we will assume that propagator
zeros are kinematical, {\it i.e.}, that bound states will form for each of the gapped
mirror fermions.   That leaves us to consider zeros of $H_{\rm eff}(\vec p)$.
$H_{\rm eff}(\vec p)$ is not analytic at such zeros, as we will see below, but it should
still have a continuous first derivative at such {\it degeneracy points} (poles
of ${\cal{R}}(\vec p)$, which, of course, yield the zeros of $H_{\rm eff}(\vec p)$).

We will now make use of our assumption that only relativistic massless
fermions appear in the continuum limit, so that $H_{\rm eff}(\vec p)$
has to take the form
\begin{eqnarray}
\label{Hzeros}
H_{\rm eff}(p)&=&\pm(p-p_c)+\dots\ ,\quad\qquad\ \ \mbox{(d=1)}\ ,\\
H_{\rm eff}(\vec p)&=&\pm{\vec{\sigma}}\cdot(\vec{p}-\vec{p}_c)+\dots\ ,\qquad \mbox{(d=3)}\ ,
\nonumber
\end{eqnarray}
where $\vec\sigma$ are the Pauli matrices, and $p_c$ is a degeneracy
point.\footnote{One can always redefine momentum such that $p_c=0$.}
For a generic dimensionless SMG coupling $g$, with the assumption that the SMG
interactions are irrelevant near the continuum limit, the lowest-order non-analytic loop contribution
to $H_{\rm eff}$ appears at two loops, taking the generic form
\begin{eqnarray}
\label{Hloops}
H_{\rm eff}(p)&=&\pm p\left(1+c_1g^2(ap)^{2(n-2)}\log(p^2)+\dots\right)\ ,\quad\qquad\ \ \mbox{(d=1)}\ ,\\
H_{\rm eff}(\vec p)&=&\pm{\vec{\sigma}}\cdot\vec{p}\left(1+c_3g^2(ap)^{2(n-4)}\log(p^2)+\dots\right)\ ,\qquad \mbox{(d=3)}\ ,\nonumber
\end{eqnarray}
where we made use of the freedom to redefine momentum such that $p_c=0$, 
and
where $n>d+1$ if indeed all interactions are irrelevant near the continuum limit;
$c_1$ and $c_3$ are numerical constants.
Thus, by inspection, $H_{\rm eff}$ has a continuous first derivative.

Thus, under the assumptions that \cite{GSNN}:
\begin{itemize}
\item The lattice hamiltonian has a finite range and contains lattice fermions and SMG
interactions only (these assumptions can be relaxed),
\item The lattice model has exactly conserved discrete charges (no SSB takes
place),
\item The continuum limit contains only massless free relavistic fermions, and no massless
bosons,
\item In each charge sector, one can construct a complete set of interpolating
fields (which implies that ${\cal{R}}(\vec p)$ has no zeros),
\end{itemize}
it then follows that $H_{\rm eff}(\vec p)$ satisfies all the conditions of the NN
theorem, and therefore the massless spectrum is vectorlike.

\section{Roadmap}
With the conclusion of the previous section, does this mean that all recent
attempts, including Ref.~\cite{ZZWY}, are doomed to fail?   Not quite!

We begin with noting all the assumptions we had to make to extend the proof
of the NN theorem to the interacting case.   There are very many potential
SMG models, and we have not tried to somehow classify all of them.  Rather,
our results imply a number of ``homework'' problems that have to be carried
out for any proposed SMG-based construction of a lattice chiral gauge theory:
\begin{itemize}
\item Can the SMG interactions
(which necessarily need to be non-perturbatively strong) generate bound states?
Are propagator zeros genuine or kinematical?
\item Are the massless fermions free in the continuum limit?   We remind the 
reader that a chiral gauge theory in which the gauge interactions are turned
off is a theory of free massless Weyl fermions.   Therefore, this is what the
reduced model has to yield in the continuum limit.

\item We note that
the model of Ref.~\cite{ZZWY} is in 2 dimensions, and that one thus expects
marginal 4-fermion operators to be induced. If indeed this happens, this model is not
free in the continuum limit, and thus does not lead to the 3450 U(1) chiral gauge theory (unless all marginal 4-fermion interactions are tuned to zero).   Of course, this
particular question is important only in 2 dimensions.

\item A good way to test the massless fermion content is to compute the
vacuum polarization.   This is the two-point function of the conserved current
to which the gauge field will couple when it is turned on.  Its singularity structure
near zero external momentum provides non-trivial information about massless
states present in the reduced model.
\end{itemize}

We end with an observation.   A lot of emphasis in the literature has been devoted
to the anomaly structure of SMG models.   And of course, any attempt that does
not take care to cancel all gauge anomalies is doomed to fail once the gauge
fields are turned on.   However, the 
key question of the applicability of the NN theorem to the reduced model is a 
question about the model without any gauge fields.   The reduced model does
not, by itself, appear to care about any anomalies!   In fact, in our analysis of the
extension of the NN theorem in the context of the SMG paradigm, anomalies
appear to play no role \cite{GSNN}.  Thus, taking care of anomaly cancellation
is no guarantee for success.\\
\bigskip

\noindent
{\bf Acknowledgments.\ }
We thank Cenke Xu and Yi-Zhuang You for fruitful discussions.
MG and YS both benefited from the workshop on Symmetric
Mass Generation held in May 2024 at the Simons Center
for Geometry and Physics at SUNY Stony Brook.
MG also thanks the Kavli Institute of Theoretical Physics at UC Santa Barbara,
where part of this work was carried out during the program
``What is Particle Theory,'' for hospitality.
This material is based upon work supported by the U.S. Department of
Energy, Office of Science, Office of High Energy Physics,
Office of Basic Energy Sciences Energy Frontier Research Centers program
under Award Number DE-SC0013682 (MG).
YS is supported by the Israel Science Foundation under grant no.~1429/21.

\end{document}